\documentclass[prl,twocolumn,showpacs,superscriptaddress,floatfix,preprintnumbers]{revtex4}
\usepackage{graphicx}
\usepackage{dcolumn}

\def\be{\begin{equation}}
\def\ee{\end{equation}}

\begin{document}


%
%
%
%
%
%

\title{Search for Randall-Sundrum Gravitons in the Diphoton Channel at CDF}
\affiliation{Institute of Physics, Academia Sinica, Taipei, Taiwan 11529, Republic of China} 
\affiliation{Argonne National Laboratory, Argonne, Illinois 60439, USA} 
\affiliation{University of Athens, 157 71 Athens, Greece} 
\affiliation{Institut de Fisica d'Altes Energies, Universitat Autonoma de Barcelona, E-08193, Bellaterra (Barcelona), Spain} 
\affiliation{Baylor University, Waco, Texas 76798, USA} 
\affiliation{Istituto Nazionale di Fisica Nucleare Bologna, $^z$University of Bologna, I-40127 Bologna, Italy} 
\affiliation{University of California, Davis, Davis, California 95616, USA} 
\affiliation{University of California, Los Angeles, Los Angeles, California 90024, USA} 
\affiliation{Instituto de Fisica de Cantabria, CSIC-University of Cantabria, 39005 Santander, Spain} 
\affiliation{Carnegie Mellon University, Pittsburgh, Pennsylvania 15213, USA} 
\affiliation{Enrico Fermi Institute, University of Chicago, Chicago, Illinois 60637, USA}
\affiliation{Comenius University, 842 48 Bratislava, Slovakia; Institute of Experimental Physics, 040 01 Kosice, Slovakia} 
\affiliation{Joint Institute for Nuclear Research, RU-141980 Dubna, Russia} 
\affiliation{Duke University, Durham, North Carolina 27708, USA} 
\affiliation{Fermi National Accelerator Laboratory, Batavia, Illinois 60510, USA} 
\affiliation{University of Florida, Gainesville, Florida 32611, USA} 
\affiliation{Laboratori Nazionali di Frascati, Istituto Nazionale di Fisica Nucleare, I-00044 Frascati, Italy} 
\affiliation{University of Geneva, CH-1211 Geneva 4, Switzerland} 
\affiliation{Glasgow University, Glasgow G12 8QQ, United Kingdom} 
\affiliation{Harvard University, Cambridge, Massachusetts 02138, USA} 
\affiliation{Division of High Energy Physics, Department of Physics, University of Helsinki and Helsinki Institute of Physics, FIN-00014, Helsinki, Finland} 
\affiliation{University of Illinois, Urbana, Illinois 61801, USA} 
\affiliation{The Johns Hopkins University, Baltimore, Maryland 21218, USA} 
\affiliation{Institut f\"{u}r Experimentelle Kernphysik, Karlsruhe Institute of Technology, D-76131 Karlsruhe, Germany} 
\affiliation{Center for High Energy Physics: Kyungpook National University, Daegu 702-701, Korea; Seoul National University, Seoul 151-742, Korea; Sungkyunkwan University, Suwon 440-746, Korea; Korea Institute of Science and Technology Information, Daejeon 305-806, Korea; Chonnam National University, Gwangju 500-757, Korea; Chonbuk National University, Jeonju 561-756, Korea} 
\affiliation{Ernest Orlando Lawrence Berkeley National Laboratory, Berkeley, California 94720, USA} 
\affiliation{University of Liverpool, Liverpool L69 7ZE, United Kingdom} 
\affiliation{University College London, London WC1E 6BT, United Kingdom} 
\affiliation{Centro de Investigaciones Energeticas Medioambientales y Tecnologicas, E-28040 Madrid, Spain} 
\affiliation{Massachusetts Institute of Technology, Cambridge, Massachusetts 02139, USA} 
\affiliation{Institute of Particle Physics: McGill University, Montr\'{e}al, Qu\'{e}bec, Canada H3A~2T8; Simon Fraser University, Burnaby, British Columbia, Canada V5A~1S6; University of Toronto, Toronto, Ontario, Canada M5S~1A7; and TRIUMF, Vancouver, British Columbia, Canada V6T~2A3} 
\affiliation{University of Michigan, Ann Arbor, Michigan 48109, USA} 
\affiliation{Michigan State University, East Lansing, Michigan 48824, USA}
\affiliation{Institution for Theoretical and Experimental Physics, ITEP, Moscow 117259, Russia}
\affiliation{University of New Mexico, Albuquerque, New Mexico 87131, USA} 
\affiliation{Northwestern University, Evanston, Illinois 60208, USA} 
\affiliation{The Ohio State University, Columbus, Ohio 43210, USA} 
\affiliation{Okayama University, Okayama 700-8530, Japan} 
\affiliation{Osaka City University, Osaka 588, Japan} 
\affiliation{University of Oxford, Oxford OX1 3RH, United Kingdom} 
\affiliation{Istituto Nazionale di Fisica Nucleare, Sezione di Padova-Trento, $^{aa}$University of Padova, I-35131 Padova, Italy} 
\affiliation{LPNHE, Universite Pierre et Marie Curie/IN2P3-CNRS, UMR7585, Paris, F-75252 France} 
\affiliation{University of Pennsylvania, Philadelphia, Pennsylvania 19104, USA}
\affiliation{Istituto Nazionale di Fisica Nucleare Pisa, $^{bb}$University of Pisa, $^{cc}$University of Siena and $^{dd}$Scuola Normale Superiore, I-56127 Pisa, Italy} 
\affiliation{University of Pittsburgh, Pittsburgh, Pennsylvania 15260, USA} 
\affiliation{Purdue University, West Lafayette, Indiana 47907, USA} 
\affiliation{University of Rochester, Rochester, New York 14627, USA} 
\affiliation{The Rockefeller University, New York, New York 10065, USA} 
\affiliation{Istituto Nazionale di Fisica Nucleare, Sezione di Roma 1, $^{ee}$Sapienza Universit\`{a} di Roma, I-00185 Roma, Italy} 

\affiliation{Rutgers University, Piscataway, New Jersey 08855, USA} 
\affiliation{Texas A\&M University, College Station, Texas 77843, USA} 
\affiliation{Istituto Nazionale di Fisica Nucleare Trieste/Udine, I-34100 Trieste, $^{ff}$University of Trieste/Udine, I-33100 Udine, Italy} 
\affiliation{University of Tsukuba, Tsukuba, Ibaraki 305, Japan} 
\affiliation{Tufts University, Medford, Massachusetts 02155, USA} 
\affiliation{University of Virginia, Charlottesville, VA  22906, USA}
\affiliation{Waseda University, Tokyo 169, Japan} 
\affiliation{Wayne State University, Detroit, Michigan 48201, USA} 
\affiliation{University of Wisconsin, Madison, Wisconsin 53706, USA} 
\affiliation{Yale University, New Haven, Connecticut 06520, USA} 
\author{T.~Aaltonen}
\affiliation{Division of High Energy Physics, Department of Physics, University of Helsinki and Helsinki Institute of Physics, FIN-00014, Helsinki, Finland}
\author{B.~\'{A}lvarez~Gonz\'{a}lez$^v$}
\affiliation{Instituto de Fisica de Cantabria, CSIC-University of Cantabria, 39005 Santander, Spain}
\author{S.~Amerio}
\affiliation{Istituto Nazionale di Fisica Nucleare, Sezione di Padova-Trento, $^{aa}$University of Padova, I-35131 Padova, Italy} 

\author{D.~Amidei}
\affiliation{University of Michigan, Ann Arbor, Michigan 48109, USA}
\author{A.~Anastassov}
\affiliation{Northwestern University, Evanston, Illinois 60208, USA}
\author{A.~Annovi}
\affiliation{Laboratori Nazionali di Frascati, Istituto Nazionale di Fisica Nucleare, I-00044 Frascati, Italy}
\author{J.~Antos}
\affiliation{Comenius University, 842 48 Bratislava, Slovakia; Institute of Experimental Physics, 040 01 Kosice, Slovakia}
\author{G.~Apollinari}
\affiliation{Fermi National Accelerator Laboratory, Batavia, Illinois 60510, USA}
\author{J.A.~Appel}
\affiliation{Fermi National Accelerator Laboratory, Batavia, Illinois 60510, USA}
\author{A.~Apresyan}
\affiliation{Purdue University, West Lafayette, Indiana 47907, USA}
% ...
\collaboration{CDF Collaboration\footnote{With visitors from $^a$University of Massachusetts Amherst, Amherst, Massachusetts 01003,
$^b$Istituto Nazionale di Fisica Nucleare, Sezione di Cagliari, 09042 Monserrato (Cagliari), Italy,
$^c$University of California Irvine, Irvine, CA  92697, 
$^d$University of California Santa Barbara, Santa Barbara, CA 93106
$^e$University of California Santa Cruz, Santa Cruz, CA  95064,
$^f$CERN,CH-1211 Geneva, Switzerland,
$^g$Cornell University, Ithaca, NY  14853, 
$^h$University of Cyprus, Nicosia CY-1678, Cyprus, 
$^i$University College Dublin, Dublin 4, Ireland,
$^j$University of Fukui, Fukui City, Fukui Prefecture, Japan 910-0017,
$^k$Universidad Iberoamericana, Mexico D.F., Mexico,
$^l$Iowa State University, Ames, IA  50011,
$^m$University of Iowa, Iowa City, IA  52242,
$^n$Kinki University, Higashi-Osaka City, Japan 577-8502,
$^o$Kansas State University, Manhattan, KS 66506,
$^p$University of Manchester, Manchester M13 9PL, England,
$^q$Queen Mary, University of London, London, E1 4NS, England,
$^r$Muons, Inc., Batavia, IL 60510,
$^s$Nagasaki Institute of Applied Science, Nagasaki, Japan, 
$^t$National Research Nuclear University, Moscow, Russia,
$^u$University of Notre Dame, Notre Dame, IN 46556,
$^v$Universidad de Oviedo, E-33007 Oviedo, Spain, 
$^w$Texas Tech University, Lubbock, TX  79609, 
$^x$Universidad Tecnica Federico Santa Maria, 110v Valparaiso, Chile,
$^y$Yarmouk University, Irbid 211-63, Jordan,
$^{gg}$On leave from J.~Stefan Institute, Ljubljana, Slovenia, 
}}
\noaffiliation

\preprint{FERMILAB-PUB-10-498-E}
\begin{abstract}

We report on a search for new particles in the diphoton channel using a data sample of $p\bar p$ collisions at $\sqrt{s} = 1.96$ TeV collected by the CDF II detector at the Fermilab Tevatron, with an integrated luminosity of 5.4 fb$^{-1}$. The diphoton invariant mass spectrum of the data agrees well with the standard model expectation. We set upper limits on the production cross section times branching ratio for the Randall-Sundrum graviton, as a function of diphoton mass. We subsequently derive lower limits on the graviton mass of 459~GeV$/c^{2}$ and 963~GeV$/c^{2}$, at the 95\% confidence level, for coupling parameters ($k/\overline{M}_{Pl}$) of 0.01 and 0.1 respectively.  
\end{abstract}



\maketitle



The large disparity between the electroweak scale and the gravity scale (the Planck scale) is known as the hierarchy problem. In the Randall-Sundrum (RS) model \cite{Randall:1999ee}, the hierarchy is generated by introducing one extra spatial dimension. The 5-dimensional space-time is bounded by two 4-dimensional subspaces (or branes, short for membranes). The standard model (SM) particles are confined to the ``TeV'' brane, located at $\phi = \pi$, while the Planck brane is located at $\phi = 0$, where $\phi$ is the angular coordinate parametrizing the extra dimension ($0\leq|\phi|\leq\pi$). Gravity is localized on the Planck brane but can propagate in the bulk. The apparent weakness of gravity arises from the small overlap of the gravitational wave function with the TeV brane. The scale of physical phenomena on the TeV brane is generated from the Planck scale through a warp factor: $\Lambda_{\pi}=\overline{M}_{Pl}e^{-kr_{c}\pi}$, where $\Lambda_{\pi} \sim$ TeV, $\overline{M}_{Pl}=M_{Pl}/\sqrt{8\pi}$ is the reduced Planck scale, $k$ is the curvature scale of the extra dimension, and $r_{c}$ is the compactification radius of the extra dimension. The hierarchy is reproduced if $kr_{c}\simeq12$.

The compactification of the extra dimension gives rise to a Kaluza-Klein (KK) tower of graviton states, the mass spectrum being $m_{n}=x_{n}(k/\overline{M}_{Pl})\Lambda_{\pi}$, where $x_{n}$ is the $n^{th}$ root of the first-order Bessel function, and the states couple with strength $1/\Lambda_{\pi}$. Two parameters determine graviton couplings and widths: the constant $k/\overline{M}_{Pl}$ and the mass of the first KK graviton excitation $m_{1}$. We examine values in the range $0.01\leq k/\overline{M}_{Pl}\leq0.1$ since the values of $k$ must be large enough to be consistent with the apparent weakness of gravity, but small enough to prevent the theory from becoming non-perturbative \cite{Davoudiasl:1999jd}. The graviton decay modes are expected to produce distinctive final states: the diphoton state from $s$ wave decays and the dilepton state from $p$ wave decays.  The spin-2 nature of the graviton favors search in the diphoton channel, where the branching ratio (4\%) is twice that of any single dilepton channel (2\%)~\cite{Han:1998sg}.


In this paper, we report on a search for the first KK graviton excitation of the RS model in the diphoton decay channel. We use 5.4~fb$^{-1}$ of integrated luminosity collected by the CDF II detector at the Fermilab Tevatron using $p\bar{p}$ collisions at $\sqrt{s} = 1.96$ TeV between February 2002 and June 2009. Existing lower mass limits on RS gravitons from the previous CDF analysis using data corresponding to 1.2~fb$^{-1}$ of integrated luminosity in the diphoton channel are 230~GeV/$c^{2}$ and 850~GeV/$c^{2}$ for $k/\overline{M}_{Pl} = 0.01$ and 0.1 respectively at 95\% confidence level (C.L.) \cite{:2007kpa}. The most recent limits from the D0 collaboration are from a combined diphoton and dielectron search using data corresponding to 5.4~fb$^{-1}$ of integrated luminosity, with limits of 560~GeV/$c^{2}$ and 1050~GeV/$c^{2}$ for $k/\overline{M}_{Pl} = 0.01$ and 0.1 respectively~\cite{Abazov:2010xh}. 

The CDF II detector has a cylindrical geometry with forward-backward and azimuthal symmetry. It consists of a tracking system in a 1.4~T magnetic field, coaxial with the beam, surrounded by calorimeters and muon detection chambers \cite{co}. The tracking system consists of a silicon tracker (SVX-II) \cite{Sill:2000zz} and an open cell drift chamber (COT) \cite{Affolder:2004}. COT covers the pseudorapidity range $|\eta|<1.0$ \cite{co}, and the silicon detector extends the tracking coverage to $|\eta|<2.0$. The central and plug calorimeters \cite{Balka:1987ty} are sampling calorimeters that surround the COT and cover the range $|\eta|<1.1$ and $1.2<|\eta|<3.6$, respectively. The calorimeters, consisting of electromagnetic (EM) and hadronic layers arranged in a projective geometry, allow measurement of the ``transverse energy'' $E_{T} = E\sin(\theta)$ \cite{co}.  At the approximate electromagnetic shower maximum, the EM calorimeters contain fine-grained detectors \cite{Apollinari:1998bg} that measure the shower shape and centroid position in the two dimensions transverse to the shower development. Surrounding these detectors is a system of muon detectors \cite{Ascoli:1987aw}. A three-level real-time event-selection system (trigger) filters events. 

The events used in this analysis are selected by at least one of four triggers. Two of them require two clusters of electromagnetic energy: one requires both clusters to have transverse energy $E_{T} > 12$~GeV and be isolated in the calorimeter; the other requires the two clusters to have $E_{T} > 18$~GeV but makes no isolation requirement. To ensure very high trigger efficiency for large $E_{T}$ photons, events are also accepted from two single photon triggers with no isolation requirements. One requires $E_{T} > 50$~GeV, while the other requires $E_{T} > 70$~GeV with relaxed requirements on the hadronic energy associated with clusters. The combination of these triggers is effectively 100\% efficient for the kinematic region used in this search for diphoton events with an invariant mass above 100~GeV/$c^{2}$. 

In the selected sample, each event is required to have at least two photon candidates. If there are more than two photon candidates, only the two photons with the highest $E_{T}$ are used. Both photons are required to be in the fiducial region of the central calorimeter (approximately in the region $|\eta|<1.04$). Each of the two photons is required to have an energy cluster predominantly in the electromagnetic calorimeter portion with $E_{T} > 15$ GeV and the photon pair is required to have a reconstructed diphoton invariant mass greater than 30~GeV/$c^{2}$. Both clusters are required to be in the fiducial region of the shower maximum detectors and to pass the following photon identification criteria: transverse shower profiles consistent with a single photon, additional transverse energy in the calorimeter in a cone of angular radius $R = \sqrt{(\Delta\phi)^{2}+(\Delta\eta)^{2}} = 0.4$ \cite{co} around the photon candidate less than 2~GeV, and the scalar sum of the transverse momentum $p_{T}$ of the tracks in the same cone less than 2~GeV/$c$. Photons are required to have isolated energy clusters in the shower maximum detector.  

\begin{figure}[htp]
\begin{center}
\includegraphics[width=1.0\linewidth]{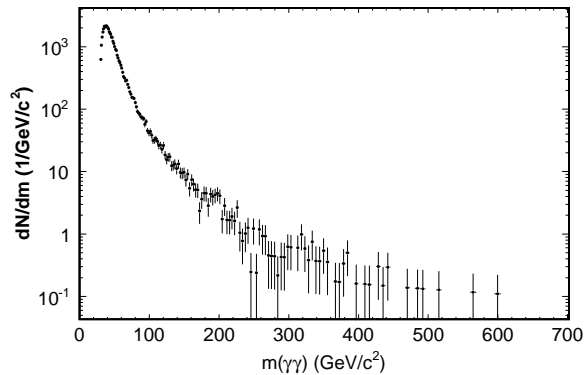}
\end{center}
\caption{The diphoton invariant mass distribution of events, histogrammed in bins of approximately one unit of calorimeter mass resolution. The width of the horizontal bar represents the bin size. The bin size is 2.4 (10.4) GeV/$c^{2}$ at $m(\gamma\gamma)$ = 100 (500) GeV/$c^{2}$.}
\label{Fig:all_mass}
\end{figure}

The selected data consist of 47 920 events. The diphoton invariant mass distribution for these events, histogrammed in bins equivalent to the mass resolution (approximated by $0.13\sqrt{m}\oplus0.02m$~GeV/$c^{2}$ \cite{Balka:1987ty}, where $m$ is diphoton mass in GeV/$c^{2}$) is shown in Fig.\ref{Fig:all_mass}. The highest mass pair occurs at 603~GeV/$c^{2}$.

The expected number of RS graviton events, as a function of graviton mass, is estimated using the {\sc pythia6.226} event generator \cite{Sjostrand:2000wi}, with CTEQ5L parton distribution functions (PDFs) \cite{Pumplin:2002vw}, and processed by the {\sc geant3}-based CDF II detector simulation \cite{geant}. The photon selection efficiency determined from simulation is multiplied by a correction factor which is derived as the ratio of the measured and simulated detector response to electrons from $Z$ boson decays, since a pure sample of reconstructed photons is not available, and the characteristics of energy deposited in the calorimeter by electrons are almost identical to those of photons. The $Z^{0}\rightarrow e^{+}e^{-}$ sample is also used to calibrate the electromagnetic energy scale. The diphoton energy scale in data and Monte Carlo is corrected by tuning the $Z^{0}\rightarrow e^{+}e^{-}$ mass peak to the world average value \cite{Amsler:2008zzb}. We also correct for effects introduced by multiple interactions in the same beam crossing. The combined acceptance and selection efficiency for RS diphoton events increases from $0.12\pm 0.01(\mbox{stat}) \pm 0.01(\mbox{syst})$ for gravitons of mass 200~GeV/$c^{2}$ to $0.33\pm 0.01(\mbox{stat}) \pm 0.03(\mbox{syst})$ for gravitons of mass 1100~GeV/$c^{2}$. The largest systematic uncertainties on the expected number of graviton events arise from the luminosity measurement (6\%) and the uncertainty associated with the initial and final state radiation (ISR/FSR). The uncertainty on the efficiency resulting from ISR/FSR is estimated to decrease from 8\% for gravitons of mass 200~GeV/$c^{2}$ to 4\% for gravitons of mass 1100~GeV/$c^{2}$ by varying the parton shower parameters in {\sc pythia}.

There are two significant background components in the diphoton data sample. The first is SM diphoton production. We estimate the shape of this background with the {\sc diphox} next-to-leading-order (NLO) Monte Carlo \cite{Binoth:1999qq} calculation. This program calculates the cross section for diphoton production in the hadronic collisions as a function of mass. Previous studies show that {\sc diphox} describes the shape of SM diphoton invariant mass spectrum well in the range used for this analysis (diphoton invariant mass above 100 GeV/$c^{2}$)~\cite{cdfdiphotonxsec}.
The mass distribution of the {\sc diphox} calculation is fitted to a product of a polynomial and the sum of five exponential distributions in the range 30 GeV/$c^{2}$ to 1.3 TeV/$c^{2}$. The fitted invariant mass spectrum is then multiplied by an efficiency function derived from a SM diphoton sample generated by {\sc pythia} and processed through the full detector simulation. The second background component, negligible except at the lowest masses, arises from the misidentification of one or two jets as photons. The invariant mass shape of this background is parameterized with a product of a polynomial and the sum of two exponentials. This functional form is justified, but not fixed, by a study using a sample of photon-like jets obtained by loosening the photon selection criteria (transverse shower profile and isolation requirements) for both photon candidates, then removing the events which pass all the signal selection requirements. 


To find the most accurate description of the background invariant mass for setting limits, we fit a functional form which is a sum of the {\sc diphox} shape and the photon-like jets shape to the invariant mass spectrum of the data. All the parameters in the functional form for the photon-like jets and the normalization of the SM diphoton background are allowed to vary in the fit.
We fit the invariant mass spectrum in the range $m_{\gamma\gamma} > 100$~GeV/$c^{2}$ since RS gravitons have been excluded at the 95\% C.L. in the lower mass region by previous searches. 

\begin{figure}[htp]
\begin{center}
\includegraphics[width=1.0\linewidth]{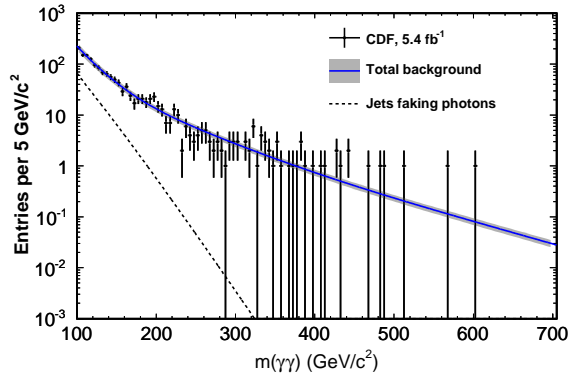}
\end{center}
\caption{
The diphoton invariant mass spectrum with the fitted total background overlaid. The points are the data. The width of the horizontal bar represents the bin size which is fixed at 5 GeV/$c^{2}$. The dotted line shows the contribution from events where at least one selected photon is from a misidentified jet, and the solid line shows this background plus the {\sc diphox} SM diphoton distribution. For clarity, the SM background is not shown in the figure since it is almost indistinguishable from the total background. The gray band shows the uncertainty on the total background.
}
\label{Fig:display_signal}
\end{figure}

Figure~\ref{Fig:display_signal} shows the observed mass spectrum with the fitted total background overlaid. The best fit SM background normalization is consistent with the {\sc diphox} calculation. The contribution of jets faking photons is highly suppressed at high diphoton masses since we restrict the amount of energy allowed in the isolation cone of each photon candidate. Also shown in the figure is the systematic uncertainty on the total background, which is approximately 20\%. The systematic uncertainty arises predominantly from the choice of the $Q^{2}$ scales used in the {\sc diphox} calculation for low mass diphoton events and from the choice of the PDFs for high mass diphoton events. The systematic uncertainties are taken to be completely correlated across all mass bins. Correlations in the systematic uncertainties between signal and background are taken into account.


A model-independent search for an excess over SM predictions is performed, following the procedure outlined in~\cite{:2007sb}. The search is optimized for a narrow resonance, but still retains sensitivity to other signals which would produce an excess over SM predictions. We scan a mass window over the mass region 100-700~GeV/$c^{2}$. The mass window is approximately the width a narrow resonance would have if observed in the CDF detector. The probability that the background could give rise to the observed number of events in the mass window, referred to as the $p$ value, is calculated using Poisson statistics.  The uncertainty on the background estimate is treated as a nuisance parameter with a Gaussian distribution. The lowest $p$ value observed is 0.016 at 198~GeV/$c^{2}$. The method is repeated for 200,000 simulated experiments produced using the background prediction. Approximately 60\% of the simulated experiments give a minimum $p$ value equal to or less than 0.016. Therefore we conclude the observed diphoton invariant mass spectrum is consistent with the background prediction.


We use  the $CL_{s}$ limit-setting technique \cite{Junk:1999kv} to set the upper limits for the production cross section of RS gravitons times the branching fraction into the $\gamma\gamma$ final state using the diphoton mass spectrum. In this method, the data are compared against two models at a time. One is the null hypothesis $H_{0}$, which asserts that the SM diphoton production and misidentified jets describe the data, while the other is the signal at a fixed mass plus background hypothesis $H_{1}$. The probability ratio
\begin{equation}
CL_{s} = \frac{P_{H_{1}}(\Delta\chi^{2}\geq\Delta\chi^{2}_{obs})}{P_{H_{0}}(\Delta\chi^{2}\geq\Delta\chi^{2}_{obs})}. 
\end{equation}
is used to set the limits. The numerator and denominator are calculated by generating simulated experiments assuming hypothesis $H_{1}$ and hypothesis $H_{0}$ respectively and taking into account systematic uncertainties on signal and background predictions. $\Delta\chi^{2}_{(obs)}$ is the difference in the logarithm likelihood values calculated by comparing the data in the simulated (real) experiment against the prediction of $H_{0}$ or $H_{1}$. The 95\% C.L. upper limit corresponds to the cross section which gives $CL_{s} = 0.05$. 

\begin{figure}[htp]
\begin{center}
\includegraphics[width=1.0\linewidth]{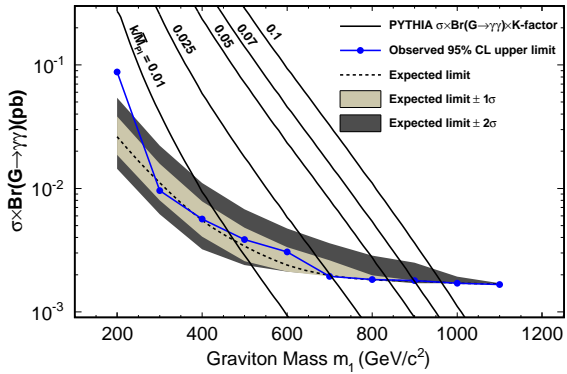}
\end{center}
\caption{The 95\% C.L. upper limit on the production cross section 
times branching fraction of an RS model graviton decaying to 
diphotons $(\sigma \times Br(G\rightarrow \gamma\gamma))$ as a 
function of graviton mass.
Also shown are the predicted $(\sigma \times Br)$ 
curves for $k/\overline{M}_{Pl}$ = 0.01, 0.07, 0.05 and 0.1~\cite{Kumar:2009nn}.}
\label{Fig:upper_limit}
\end{figure}

\begin{figure}[htp]
\begin{center}
\includegraphics[width=1.0\linewidth]{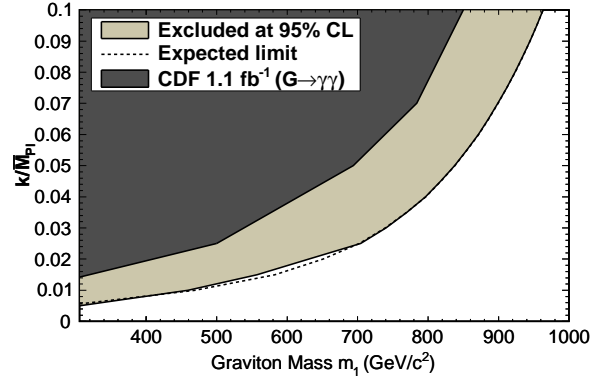}
\end{center}
\caption{The 95\% C.L. excluded region 
in the plane of $k/\overline{M}_{Pl}$ and graviton mass from 5.4 fb$^{-1}$ of integrated luminosity compared with the expected limit and the previously published exclusion contour~\cite{:2007kpa}. 
}
\label{Fig:exclude_region}
\end{figure}

The result is shown in Fig.~\ref{Fig:upper_limit}, as a function of graviton mass, along with the theoretical cross section times branching ratio for RS gravitons with $k/\overline{M}_{Pl}$ set to 0.1, 0.07, 0.05, 0.025 and 0.01. The leading-order graviton production cross section calculated by {\sc pythia}~\cite{Sjostrand:2000wi} is multiplied by a K-factor \cite{Kumar:2009nn}, decreasing from 1.54 at 200~GeV/$c^{2}$ to 0.95 at 1100~GeV/$c^{2}$, to correct for diagrams at higher-order in $\alpha_{s}$. The previous analysis \cite{:2007kpa} used a mass-independent K-factor of 1.3, which leads to conservative limits at low masses and optimistic limits at high masses. From the limit on $\sigma\times Br(G\rightarrow\gamma\gamma)$, lower mass bounds are derived for the first excited state of the RS graviton as a function of the parameter $k/\overline{M}_{Pl}$. The 95\% C.L. excluded region in the $k/\overline{M}_{Pl}$ and graviton mass plane is displayed in Fig.~\ref{Fig:exclude_region}, with the mass limits summarized in Table~\ref{Tbl:mass_limits}.

\begin{center}
\begin{table}
\begin{center}
\caption{The 95\% C.L. lower limits on the mass of the first excited state of the RS graviton for the specified values of $k/\overline{M}_{Pl}$.}
\begin{tabular}{c c} \hline\hline
$k/\overline{M}_{Pl}$ & Lower Mass Limit (GeV$/c^2$)\\ \hline
0.1 & 963\\
0.07 & 899\\
0.05 & 838\\
0.025 & 704\\
0.01 & 459\\ \hline\hline
\end{tabular}
\label{Tbl:mass_limits}
\end{center}
\end{table}
\end{center}

In conclusion, we have searched for evidence of an anomalous peak in the diphoton mass spectrum using data corresponding to an integrated luminosity of 5.4~fb$^{-1}$ collected by the CDF II detector at the Fermilab Tevatron. We find no evidence of new physics. We evaluate one model of hypothetical new diphoton production and exclude RS gravitons below masses ranging from 459 to 963~GeV/$c^{2}$, for a coupling parameter $k/\overline{M}_{Pl}$ of 0.01 to 0.1, at the 95\% C.L.. This results in a significant improvement, at high mass, over the previous best available limit in diphoton state from CDF. The limits are less stringent than those from D0 using the same integrated luminosity \cite{Abazov:2010xh}. Some contributing factors include the omission of the dielectron data and the use of mass-dependent K-factors in this CDF analysis.

We thank the Fermilab staff and the technical staffs of the participating institutions for their vital contributions. This work was supported by the U.S. Department of Energy and National Science Foundation; the Italian Istituto Nazionale di Fisica Nucleare; the Ministry of Education, Culture, Sports, Science and Technology of Japan; the Natural Sciences and Engineering Research Council of Canada; the National Science Council of the Republic of China; the Swiss National Science Foundation; the A.P. Sloan Foundation; the Bundesministerium f\"ur Bildung und Forschung, Germany; the World Class University Program, the National Research Foundation of Korea; the Science and Technology Facilities Council and the Royal Society, UK; the Institut National de Physique Nucleaire et Physique des Particules/CNRS; the Russian Foundation for Basic Research; the Ministerio de Ciencia e Innovaci\'{o}n, and Programa Consolider-Ingenio 2010, Spain; the Slovak R\&D Agency; and the Academy of Finland. We thank M.C.~Kumar, P.~Mathews, V.~Ravindran, and A.~Tripathi for the calculation of NLO K-factors for this analysis.




\begin{thebibliography}{99.}
\bibitem{co}
 A cylindrical coordinate system, $(r,\phi,z)$ is used with origin at the geometric center of the detector. $r$ is the radius from the nominal beam line, $\phi$ is the azimuthal angle, and $+z$ points along the direction of the proton beam. The polar angle $\theta$ with respect to the proton beam defines the pseudorapidity $\eta$ which is given by $\eta = - \ln[\tan(\theta/2)]$. Transverse energy and transverse momentum are defined as $E_{T}=E\sin(\theta)$ and $p_{T}=p\sin(\theta)$, respectively.
\end{thebibliography}
\end{document}